\documentclass[aps,prd,twocolumn,superscriptaddress]{revtex4}

\usepackage{bm}
\usepackage{amsfonts,amssymb}
\usepackage{amsmath}
\usepackage{graphicx,graphics}
\usepackage{enumerate}
\usepackage{subfigure}
\usepackage{appendix}
\usepackage{siunitx}
\usepackage{mathrsfs}  
\usepackage{hyperref}
\usepackage[normalem]{ulem}

\DeclareMathOperator{\csch}{csch}

\newcommand{\mm}{\Omega}
\DeclareMathOperator{\ads}{AdS}

\begin{document}

\title{Analogue gravity and radial fluid flows: The case of AdS and its deformations}

\author{David Q. Aruquipa}
\email{dquispe@ifi.unicamp.br}
\affiliation{IFGW, Universidade Estadual de Campinas, 13083-859, Campinas, SP, Brazil}
\author{Ricardo A. Mosna}
\email{mosna@ime.unicamp.br}
\affiliation{Departamento de Matem\'atica Aplicada, Universidade Estadual de Campinas, 13083-859, Campinas, SP, Brazil}
\author{Jo\~ao Paulo M. Pitelli}
\email{pitelli@ime.unicamp.br}
\affiliation{Departamento de Matem\'atica Aplicada, Universidade Estadual de Campinas, 13083-859, Campinas, SP, Brazil}

\begin{abstract}
An analogue model for the $\ads_2$ spacetime has been recently introduced by Mosna, Pitelli and Richartz [Phys. Rev. D {\bf 94}, 104065 (2016)] by considering sound waves propagating on a fluid with an ill-defined velocity profile at its source/sink. The wave propagation is then uniquely defined only when one imposes an extra boundary condition at the source/sink (which corresponds to the spatial infinity of $\ads_2$). Here we show that, once this velocity profile is smoothed out at the source/sink, the need for extra boundary conditions disappears. This, in turn, corresponds to deformations of the $\ads_2$ spacetime near its spatial infinity.  We also examine how this regularization of the velocity profile picks up a specific boundary condition for the idealized system, so that both models agree in the long wavelength limit.
\end{abstract}

\maketitle

\section{Introduction}
\label{sec:intro}

It is well known  \cite{unruh,barcelo} that sound waves propagating on a moving (inviscid, irrotational and barotropic) fluid satisfy the wave equation on an effective curved spacetime,
\begin{equation}
\frac{1}{\sqrt{-g}}\partial_\mu\left(\sqrt{-g}g^{\mu\nu}\partial_{\nu}\psi\right)=0,
\label{wave equation}
\end{equation}
where $\psi$ is the perturbation of the velocity potential  of the fluid, i.e., $\delta\vec{v}=-\vec{\nabla}\psi$. The metric $g_{\mu\nu}$ is the effective metric felt by the sound waves and it is determined by the essential quantities which define  the background fluid. In two-dimensional models this metric is given by~\cite{barcelo}
\begin{equation}
ds^2=\left(\frac{\sigma}{c}\right)^2\left[-\left(c^2-v^2\right)dt^2-2vdtdr+dr^2+r^2d\theta^2\right],
\label{metric}
\end{equation}
where $v$ is the velocity of the background fluid, $\sigma$ its density and $c$ the speed of sound relative to the fluid. In this way, sound waves propagating in a moving fluid can be used as a playground to study aspects of general relativity (analogue gravity).

Analogue models are useful to test elements of quantum field theory (QFT) in curved spaces. The detection of Hawking radiation in a fluid analogue gravity system has been reported in~\cite{Weinfurtner} (it has been argued in \cite{Michel,Euve} that the Planckianity of the spectrum is lost under the experimental setup of~\cite{Weinfurtner}; at any rate, with the increase of Froude's number, the Planckian spectrum should be observed~\cite{Michel}). The phenomenon of superradiance has also been observed in the laboratory in this context~\cite{Torres}. On the other hand, models based on Bose-Einstein condensates have also been used to provide a description of QFT in curved spaces. Analogues of cosmological particle creation by an expanding universe~\cite{Schutzhold} and Hawking radiation~\cite{Steinhauer}  have also been observed  in the laboratory within this setup.

One difficulty faced by QFT in curved spacetimes is the question of the well-posedness of the wave equation. When the spacetime fails to be globally hyperbolic, it may be possible to have an infinite number of acceptable physical evolutions for the propagating wave~\cite{wald1,wald2,ishibashi}. These  solutions are in one-to-one correspondence with self-adjoint extensions of the spatial part of the wave operator which, in turn, correspond to extra conditions that should be imposed at the spatial boundary.

In a previous paper~\cite{anads}, some of the authors introduced an analogue model based on the anti-de Sitter ($\ads$) space in terms of a planar radial flow with a point source/sink. The $\ads$ spacetime is nonglobally hyperbolic and, as a result, it is impossible to uniquely solve the wave equation  without specifying additional boundary equations at infinity (in a certain sense, information can flow in from infinity on $\ads$). On the analogue model end, the background flow on which the sound waves propagate has constant radial velocity,
\begin{equation}
\vec{v}(\vec{r})=\alpha c\, \vec{e}_r,
\label{constant radial velocity}
\end{equation}
with constant $-1<\alpha<1$, and the spacelike infinity of $\ads$ is mapped to point source at $r=0$. Notice that $|\alpha|<1$ and thus the fluid velocity is always subsonic and no dumb holes appear in this case.

Introducing a new time coordinate given by (see~\cite{anads})
\begin{equation}
\tau=t+\frac{\alpha}{c(1-\alpha^2)}r,
\label{trans. variables}
\end{equation}
and considering circularly symmetric sound waves $\psi(t,r)$, we obtain the wave equation
\begin{equation}
\frac{\partial^2 \psi}{\partial \tau^2}=c^2(1-\alpha^2)^2\frac{\partial^2\psi}{\partial r^2}.
\label{string equation}
\end{equation}
Equation (\ref{string equation}) resembles the equation of a semi-infinite string, for which boundary conditions at $r=0$ have the usual interpretation. The need for extra boundary conditions at $\ads$ can thus be interpreted as the natural requirement of specifying boundary conditions for the sound waves at the fluid source at $r=0$. In 
\cite{anads} some of us calculated, among other things, how physical quantities, like the phase difference between ingoing and outgoing scattered waves $\delta(\omega)$, relate to those boundary conditions. We also analyzed the linear stability of the fluid configuration with relation to the chosen boundary condition.

The velocity flow considered in Eq.~(\ref{constant radial velocity})  is clearly not well defined at the source. A natural question that comes to mind is, what happens to the previous analysis if we regularize $\vec{v}(\vec{r})$ near the origin so that the flow is well defined and continuous there~\cite{endnote1}? How does the  regularization affect $\delta(\omega)$ and  what is its effect on the $\ads$ counterpart of this analogue model? The aim of this paper is to provide answers to these questions.

\section{$\ads$ analogue model}
\label{AdS analogue model}
We start by briefly reviewing the $\ads$ case; more details can be found in \cite{anads}. The continuity equation for a stationary flow with constant radial velocity~(\ref{constant radial velocity}) leads to a density for the background fluid of the form
\begin{equation}
\sigma(r)=\frac{k}{\alpha c r},
\label{density_AdS}
\end{equation}
where $k$ is a constant. Substituting Eqs.~(\ref{constant radial velocity}) and~(\ref{density_AdS}) into the line element~(\ref{metric}) and making use of the transformation of variables~(\ref{trans. variables}) we get
\begin{equation}
ds^2=\frac{k^2}{\alpha^2 c^4r^2}\left[-c^2(1-\alpha^2)d\tau^2+\frac{dr^2}{1-\alpha^2}+r^2d\theta^2\right].
\end{equation}
If we define $\bar{\tau}=c\sqrt{1-\alpha^2}\tau$ and $\bar{r}=r/\sqrt{1-\alpha^2}$ we arrive at
\begin{equation}
ds^2=\frac{1}{H^2\bar{r}^2}\left[-d\bar{\tau}^2+d\bar{r}^2\right]+\frac{(1-\alpha^2)}{H^2}d\theta^2,
\end{equation}
with $H^2=\alpha^2c^4(1-\alpha^2)/k^2$. But this is just the product of the $\ads_2$ metric in Poincar\'e coordinates with $S^1$. Therefore, circularly symmetric sound waves propagating on this fluid provide an analogue model for scalar waves on $\ads_2$~\cite{endnote2}.

Radial sound waves on this background satisfy Eq.~(\ref{string equation}), where $r>0$ is the radial coordinate. It is well known that, for waves propagating on the half-line, a boundary condition is necessary at $r=0$. The boundary condition which is required in order that the spatial part of the wave operator becomes self-adjoint  is given by the mixed boundary condition
\begin{equation}
\psi(\tau,0)+\beta\frac{\partial \psi(\tau,0)}{\partial r}=0,
\label{robin}
\end{equation}
where $\beta\in\mathbb{R}\cup\{\pm\infty\}$ is a parameter. The Dirichlet and Neumann boundary conditions correspond to $\beta=0$ and $\beta=\pm\infty$, respectively.

It was shown in~\cite{anads} that the boundary condition appears as an observable in the scattering of circularly symmetric waves; i.e., the phase difference between the ingoing and outgoing waves depends on the chosen boundary condition. The scattering solution of Eq.~(\ref{string equation}) with boundary condition~(\ref{robin}) is 
\begin{equation}
\psi(t,r)\sim \left(e^{-i\frac{\omega}{c(1-\alpha)}r}+e^{i\frac{\omega}{c(1+\alpha)}r+i\delta(z)}\right)e^{-i\omega t},
\end{equation}
where $z\equiv \frac{\beta\omega}{c(1-\alpha^2)}$ and $\delta(z)$ is given by
\begin{equation}
e^{i\delta(z)}=\frac{iz-1}{iz+1}.
\label{argument0}
\end{equation}
The boundary condition parameter $\beta$ is, therefore, encoded in the phase difference between the incoming and outgoing waves.

The linear stability of the fluid configuration also depends on the boundary condition. In particular, if $\beta>0$, there are modes of the form
\begin{equation}
\psi(t,r)\sim e^{-(1-\alpha)r/\beta}e^{c(1-\alpha^2)t/\beta}. 
\end{equation}
These modes grow exponentially in time, leading to a linear instability of the configuration. For $\beta\leq 0$ and $\beta\neq\pm\infty$, the fluid configuration is mode stable.  

\section{Regularization of the fluid velocity}

We now consider sound waves in more general two-dimensional radial fluid flows. As discussed above, we are interested in the case where the velocity can be written as 
\begin{displaymath}
\vec{v}(r)=v(r)\vec{e}_r.
\end{displaymath}
The acoustic metric for the  fluid flow, which we assume to be ideal and barotropic, then takes the form
\begin{displaymath}
g_{\mu\nu}=\left(\frac{\sigma(r)}{c}\right)^2\left(\begin{array}{ccccc}
-(c^2-v(r)^2)&-v(r)&0\\
-v(r)&1&0\\
0&0&r^2
\end{array}\right),
\end{displaymath}
where, as a result of the continuity equation, the counterpart of Eq. (\ref{density_AdS}) is given by
\begin{equation}
\sigma(r)=\frac{k}{rv(r)},
\label{density}
\end{equation}
where $k$ is a constant. Note that density diverges as $r\to0$ in all models of this kind, with finite fluid velocities. This might be circumvented by adding a $\theta$ or $z$ component to the velocity, but we do not consider this here. 

According to Eq.~(\ref{wave equation}), it is straightforward to write down the wave equation for $\psi$ with $g_{\mu\nu}$ defined by Eq.~(\ref{metric}). Before doing that, we take advantage of the fact that the acoustic metric is static  to construct the transformations
\begin{equation}
\begin{aligned}
&d\tau=dt+\frac{v}{c^2-v^2}dr,\\
&d\rho=dr,\\
&d\phi=d\theta.
\end{aligned} 
\end{equation}
The acoustic metric in these coordinates is then given by
\begin{displaymath}
g_{\mu'\nu'}=\left(\frac{\sigma}{c}\right)^2\left(\begin{array}{ccccc}
-(c^2-v(\rho)^2)&0&0\\
0&\frac{c^2}{c^2-v(\rho)^2}&0\\
0&0&\rho^2
\end{array}\right),
\end{displaymath}
and the wave equation for the perturbation $\psi(\tau,\rho,\phi)$ becomes
\begin{equation}\begin{aligned}
\frac{\partial^2\psi}{\partial \tau^2}=&c^2{\left(1-\frac{v^2}{c^2}\right)}^2\frac{\partial^2\psi}{\partial \rho^2}-c^2\frac{1-\frac{v^4}{c^4}}{v}\frac{dv}{d\rho}\frac{\partial \psi}{\partial \rho}\\
&-\frac{c^2-v^2}{\rho^2}\frac{\partial^2\psi}{\partial \phi^2}.
\label{wave prime}
\end{aligned}
\end{equation}

We now consider regularizations of the profile velocity near the source/sink. Note that in the $\ads$ analogue model, $\vec{v}(\vec{r})$ is not well defined at the origin. We thus modify it so that $\vec{v}(\vec{r})$ is at least continuous there~\cite{endnote1}. Therefore, we consider regularizations of $\vec{v}$ for which $v(\rho)$ is $0$ at $\rho=0$. For $\rho > 0$, the velocity should then increase until it becomes constant, $v=\alpha c$. The transition of $v(\rho)$ from $v(0) = 0$ to $v(\rho) = \alpha c$ is, for now, left arbitrary.  

\begin{figure}
\includegraphics[width=0.9\linewidth]{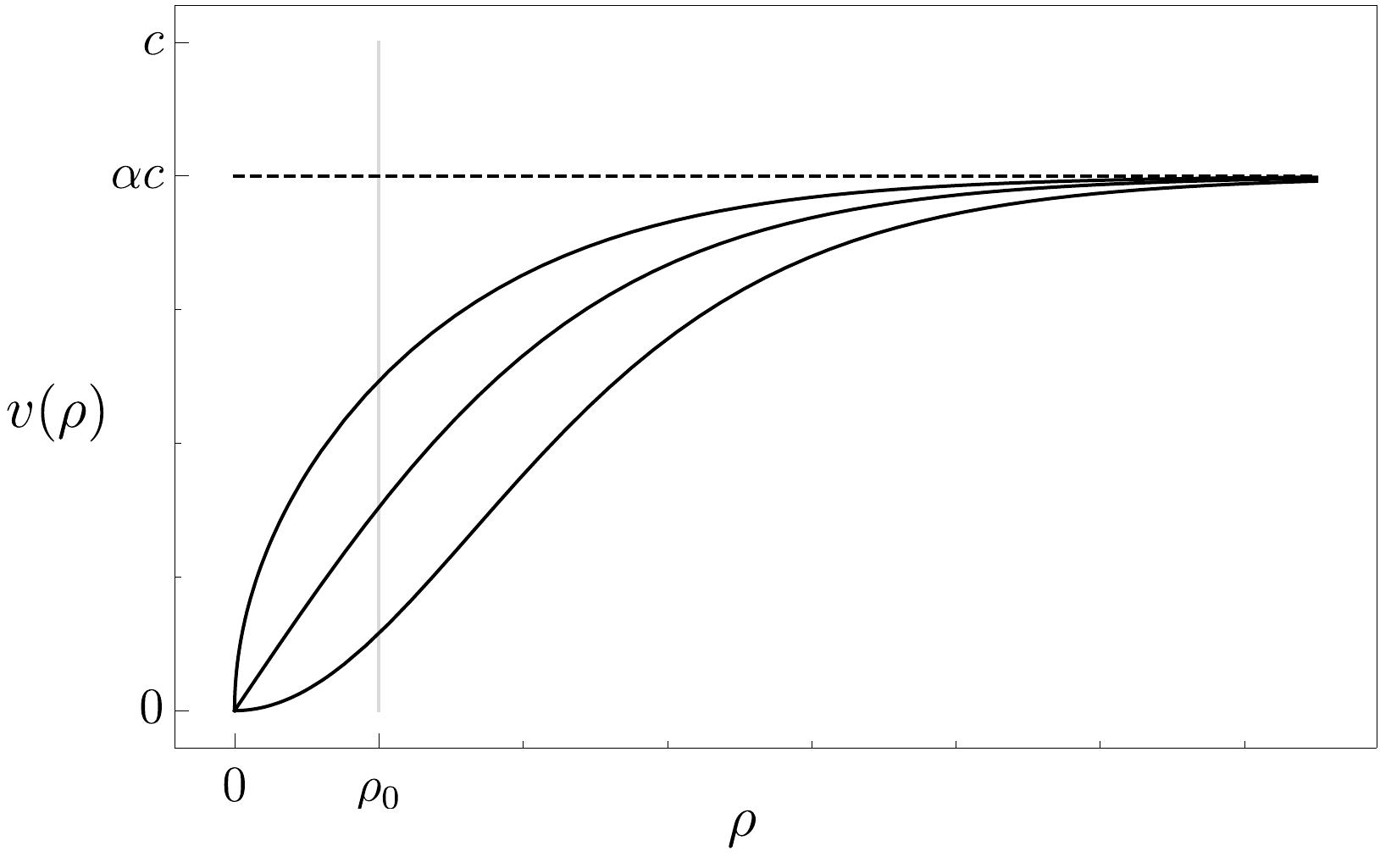}
\caption{Regularizations of $v(\rho)$ as in Eq.~(\ref{v_reg}), with $n=1/2$, $1$, and $2$ from top to bottom, respectively.}
\label{fig1}
\end{figure}

Consider the case when $v(\rho)$ can be written, near the origin, as
\begin{equation}
v(\rho)=\alpha c\left(\frac{\rho}{\rho_0}\right)^{n};\,\,\, \rho \lesssim \rho_0,
\label{v_reg}
\end{equation}
where $\rho_0$ and $n$ are undetermined parameters. This profile must still be matched to another expression, valid for $\rho> \rho_0$, which approaches the constant value $\alpha c$ for large $\rho$. We see that $\rho_0$ is related with the width of the region wherein $v(\rho)$ is not constant, and that $n$ determines how fast $v(\rho)$ grows near the origin. Figure~\ref{fig1} shows different velocity profiles corresponding to $n=1/2$, $1$, and $2$. 

We consider waves with circular symmetry and write, after separating variables, $\psi=R(\rho)e^{-i \omega \tau}$. Moreover, let us introduce the dimensionless radial coordinate
$$x\equiv\rho/\rho_0$$
and the dimensionless frequency
$$\mm\equiv\omega \rho_0/c.$$
It follows from Eq.~(\ref{wave prime}) that, near the origin (at $x=0$),
\begin{equation}
\left(1-\alpha^2x^{2n}\right)^2R''(x)-\frac{n}{x}\left(1-\alpha^4x^{4n}\right)R'(x)+\mm^2R(x)=0.
\label{Eq. for x}
\end{equation}

This equation is only meaningful for  $x\lesssim1$. We note that $x=0$ is a regular singular point of this ordinary differential equation. One solution can thus be written as 
\begin{equation}
R(x)=x^s\sum_{k=0}^{\infty}{a_kx^k},
\end{equation}
with $s=n+1$ or $s=0$ (by Frobenius method). If $n$ is noninteger, we have two linear independent solutions which, around $x=0$, behave as 
\begin{equation}
\begin{aligned}
&R_1(x)\sim x^{n+1},\\
&R_2(x)\sim 1.
\end{aligned}
\end{equation}
If $n$ is integer we have
\begin{equation}
\begin{aligned}
&R_1(x)\sim x^{n+1},\\
&R_2(x)\sim 1+ p R_1(x)\ln{x},
\end{aligned}
\end{equation}
with constant $p$. For $n=1$ (a case which is important later), we have
\begin{equation}
\begin{aligned}
&R_1(x)\sim x^{2}+\mathcal{O}(x^4),\\
&R_2(x)\sim 1-\frac{1}{2}\mm^2x^2\ln{x}+\mathcal{O}(x^4).
\end{aligned}
\label{square integrable n=1}
\end{equation}

\section{Finite energy condition}

The propagating scalar field associated with the sound waves has the energy-momentum tensor
\begin{equation}
T^{\mu\nu}=\partial^{\mu}\psi\partial^{\nu}\psi-g^{\mu\nu}\left(\frac{1}{2}g^{\alpha\beta}\partial_\alpha\psi\partial_{\beta}\psi\right).
\end{equation}
In $(\tau,\rho,\phi)$ coordinates the vector field $\xi_0^{\mu'}=\delta^{\mu'}_0$ is clearly Killing. The conserved current in this case is then given by
\begin{equation}\begin{aligned}
\mathcal{Q}^{\mu'}&=-\sqrt{-g'}T^{\mu'\nu'}\xi_\nu'=\left(\frac{\sigma^3(\rho)}{c^2}\rho\right)T^{\mu'\alpha'}g_{\alpha' 0'}\\
&= \left(\frac{\sigma^5(\rho)}{c^4}\rho\right)(c^2-v(\rho)^2)T^{\mu' 0'}.
\end{aligned}
\end{equation}

This is the conserved current in $(\tau,\rho,\phi)$ coordinates. Going to the laboratory coordinates $(t,r,\theta)$ and noticing that $\det(\partial x^{\mu'}\!\!/\partial x^{\mu})=1$, we find that the sound energy in the laboratory frame is
\begin{equation}
\mathcal{Q}^{0}=\frac{\partial x^{0}}{\partial x^{\mu'}}\mathcal{Q}^{\mu'}=\mathcal{Q}^{0'}-\frac{v(r)}{c^2-v^2(r)}\mathcal{Q}^{1}.
\label{lab frame}
\end{equation}
This energy coincides, indeed, with the usual energy density defined in the fluid dynamics literature ($\mathcal{Q}^{0}$ is obviously not invariant and, in other coordinates, it does not correspond to the correct fluid mechanics energy) \cite{stone1,stone2}.
A straightforward calculation yields
\begin{equation}\begin{aligned}
\mathcal{Q}^{0'}=&\frac{1}{2}\frac{\sigma(\rho)}{c}\rho\left[\frac{(\partial_r\psi)^2}{c^2-v(\rho)^2}+\frac{c^2-v(\rho)^2}{c^2}(\partial_r\psi)^2+\frac{(\partial_\theta\psi)^2}{\rho^2}\right]
\end{aligned}
\end{equation}
and
\begin{equation}
\mathcal{Q}^{1'}=-\frac{\sigma(\rho)}{c}\rho\frac{c^2-v(\rho)^2}{c^2}\partial_r\psi\partial_\tau\psi.
\end{equation}
The energy density in the laboratory frame is thus given by
\begin{equation}\begin{aligned}
E\equiv Q^{0}=&\frac{1}{2}\left(\frac{\sigma(r)}{c}r\right)\left\{\frac{1}{c^2}(\partial_t\psi)^2+\frac{(c^2-v(r)^2)}{c^2}(\partial_r\psi)^2\right.\\&\left.+\frac{(\partial_\theta\psi)^2}{r^2}\right\}.
\end{aligned}
\end{equation}

A straightforward calculation shows that the solutions $R_i(x)$ have energy densities $E_i$ given by
\begin{equation}\begin{aligned}
&E_1\sim \frac{kc\mm^5}{2\alpha}\sin{\left[\omega\left(t-\frac{\alpha\rho_0}{c(n+1)}x^{n+1}\right)\right]}^2 x^{2+n},\\
&E_2 \sim \frac{kc\mm^5}{2\alpha}\sin{\left[\omega\left(t-\frac{\alpha\rho_0}{c(n+1)}x^{n+1}\right)\right]}^2 x^{-n},
\end{aligned}
\end{equation}
for $x$ small (the above expressions are valid for both $n$ integer and noninteger).

For $0<n<1$, both energies are integrable and finite near $x=0$. On the other hand, for $n\geq 1$, only one of the solutions, namely, $R_1(x)$ has finite energy near $x=0$ and $R_2(x)$ must be discarded. Therefore, for $n\geq 1$ no extra boundary condition at the origin is necessary in order to reduce the number of independent solutions. For $0<n<1$ the problem is ill posed unless an extra condition is specified at $x=0$. It is interesting to note that the velocity profile is smooth at the origin precisely for $n\geq 1$. In other words, when the velocity profile is ``nice enough'' there is no need to consider self-adjoint extensions for the wave operator.

\section{An analytic solution}

To extend the previous analysis to the whole space one needs a global solution of Eq.~(\ref{wave prime}). This can be done numerically for any reasonable profile of the kind shown in Fig.~\ref{fig1}. Fluid flows that allow a closed form, exact, solution for the wave equation are hard to find but do exist. One example is given by the velocity profile given by
\begin{equation}
v(\rho)=\alpha c \tanh\left(\frac{\rho}{\rho_0}\right).
\label{tanh}
\end{equation}
Note that for $\rho\sim 0$ we have $v\sim \rho$ and therefore no boundary conditions are necessary at the origin (this is the case $n=1$ of the previous section). The radial part of the wave equation becomes, for this choice of $v$,
\begin{multline}
\left(1-\alpha^2\tanh^2{x}\right)^2\frac{d^2R(x)}{dx^2}- \\ -\frac{1-\alpha^4\tanh^4{x}}{\sinh{x}\cosh{x}}\frac{dR(x)}{dx}
+\mm^2R(x)=0,
\end{multline}
with solutions
\begin{equation}
R_{\pm}(x)=\exp{\left[\tfrac{i \mm \kappa_{\pm}(x)}{1-\alpha^2}\right]} \, 
{}_2F_1\left(a_{\mp}+\tfrac{1}{2},a_{\mp}-\tfrac{1}{2},c_{\mp},-\tfrac{\csch^2{x}}{1-\alpha^2}\right),
\end{equation}
where ${}_2F_1$ is the (ordinary) hypergeometric function. The functions $\kappa_{\pm}(x)$ and the parameters $a_{\pm}$ and $c_{\pm}$ are given by
\begin{equation}
\begin{aligned}
&\kappa_{\pm}(x)=\pm\ln{\sinh x}-\tfrac{\alpha}{2}\ln{\left(1-\alpha^2+\csch^2{x}\right)},\\
&a_{\pm}=\frac{1}{2}\pm\frac{i\mm/2}{1\pm \alpha},\\
&c_{\pm}=1\pm\frac{i\mm}{1-\alpha^2}.
\end{aligned}
\end{equation}
As discussed in the previous section, the finite energy solution $R(x)=A R_{-}(x)+B R_{+}(x)$ must be proportional to $R_1(x)$. It follows from Eq.~(\ref{square integrable n=1}) (since we are in the case $n=1$) that
\begin{equation}
A R_{-}(0)+B R_{+}(0)=0.
\end{equation}
This leads to (up to a global multiplicative constant)
\begin{equation}\begin{aligned}
&A=R_{+}(0)=\tfrac{\Gamma\left(1-\tfrac{i\mm}{1-\alpha}\right)}{\Gamma\left(1-\tfrac{i\mm/2}{1-\alpha}\right)\Gamma\left(1-\tfrac{i\mm/2}{1+\alpha}\right)} \, e^{-i\frac{\mm\ln{(1-\alpha^2)}}{2(1-\alpha)}},\\
&B=-R_{-}(0)=-\tfrac{\Gamma\left(1+\tfrac{i\mm}{1-\alpha}\right)}{\Gamma\left(1+\tfrac{i\mm/2}{1-\alpha}\right)\Gamma\left(1+\tfrac{i\mm/2}{1+\alpha}\right)} \, e^{i\frac{\mm\ln{(1-\alpha^2)}}{2(1+\alpha)}}.
\end{aligned}
\end{equation}

These expressions for $A$ and $B$ now provide a complete description for the circular waves propagating on the fluid. In particular, the way by which the waves interact with the source/sink located at $x=0$ (or equivalently $r=0$) is also encoded by these expressions.

We now find the phase difference $\delta(\mm)$ between incoming and outgoing circular waves.
For $x\gg1$, the velocity is constant and we know that, for this case, the solutions must be incoming and outgoing waves of the form $e^{\pm i \mm x/(1-\alpha^2)}$ \cite{anads}. Let us recover this behavior from the asymptotic expansions of $R_{\pm}(x)$. For the hypergeometric function, we have, for $x\to\infty$,
\begin{equation}
{}_2F_1\left(a_{\mp}+\tfrac{1}{2},a_{\mp}-\tfrac{1}{2},c_{\mp},-\tfrac{\csch^2{x}}{1-\alpha^2}\right)\sim
1+\frac{\left(1-4 a_{\mp}^2\right) }{\left(1-\alpha ^2\right) c_{\mp}} \, e^{- x}.
\end{equation}
Next, we compute the asymptotic behaviour of $\kappa_{\pm}(x)$. We have
\begin{equation}
\kappa_{\pm}(x)\sim
\pm\left(x-\ln{2}\right) -\tfrac{\alpha}{2}\ln{\left(1-\alpha^2+4 e^{-2 x}\right)}.
\end{equation}

As a result, the solution of the wave equation for $x\to\infty$ becomes, up to a multiplicative constant,
\begin{equation}
\psi(\tau,x)\sim \left( \eta \, e^{-\frac{i\mm x}{1-\alpha^2}} +\xi \, e^{\frac{i\mm x}{1-\alpha^2}}\right)e^{-i\omega\tau},
\end{equation}
where $\eta=C_{+}\, A$ and $\xi=C_{-}\, B$, with 
\begin{equation}
C_{\pm} = e^{\pm i \frac{\mm \ln{2}}{1-\alpha^2}}.
\end{equation}

The phase difference $\delta(\mm)$ between incoming and outgoing waves is thus given by
\begin{equation}
\begin{aligned}
e^{i \delta(\mm)}&=\frac{\xi}{\eta}=-\frac{\Gamma\left(1+\frac{i\mm}{1-\alpha^2}\right)\Gamma\left(1-\frac{i\mm/2}{1-\alpha}\right)\Gamma\left(1-\frac{i\mm/2}{1+\alpha}\right)}{\Gamma\left(1-\frac{i\mm}{1-\alpha^2}\right)\Gamma\left(1+\frac{i\mm/2}{1-\alpha}\right)\Gamma\left(1+\frac{i\mm/2}{1+\alpha}\right)} \times \\
&\times
\exp{\left[i\frac{\mm \ln{\left(\frac{1-\alpha^2}{4}\right)}} {1-\alpha^2} \right]}.
\end{aligned}
\label{argument}
\end{equation}

\begin{figure}
\includegraphics[width=0.9\linewidth]{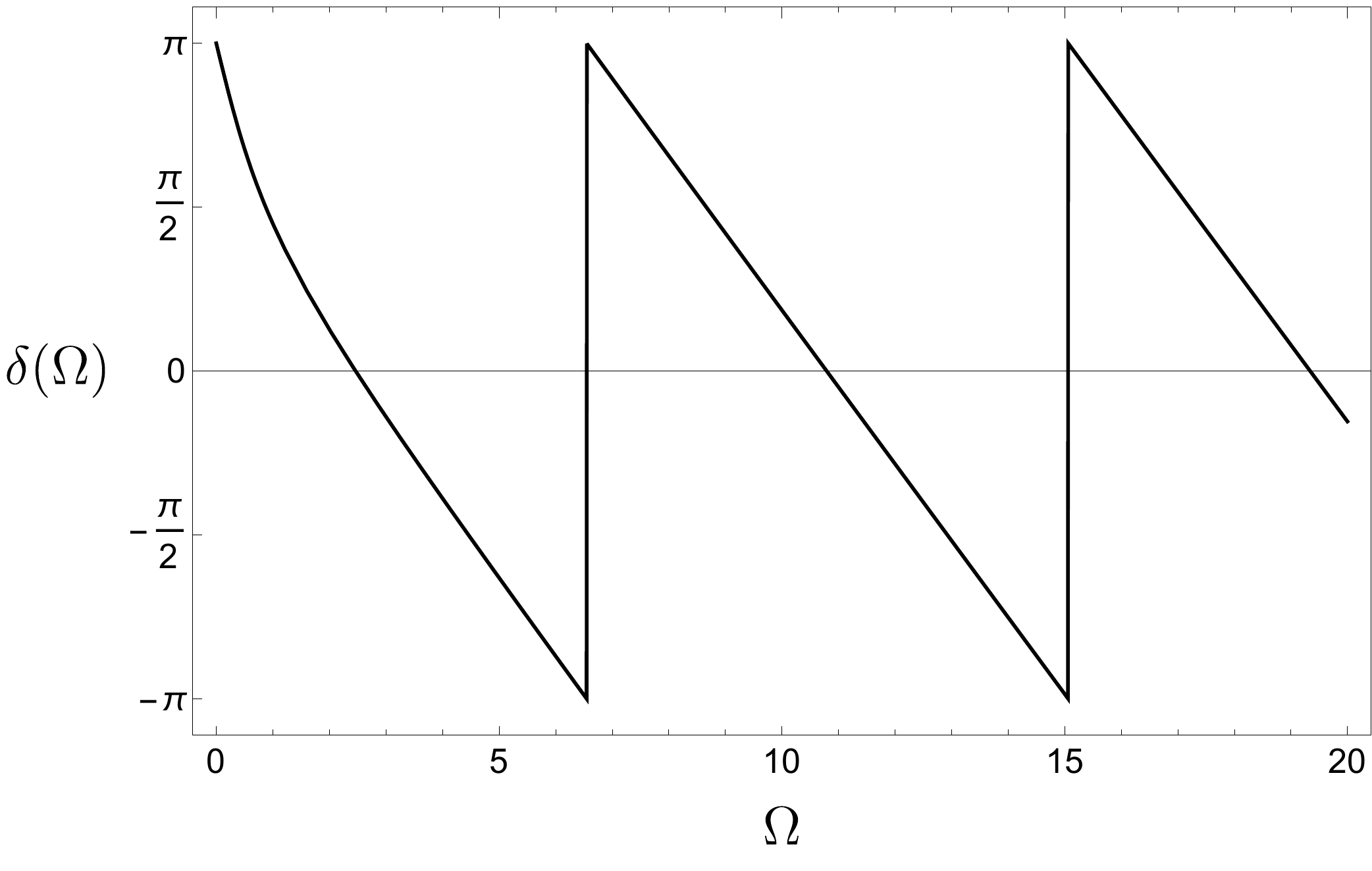}
\caption{Phase difference between ingoing and outgoing circular waves for $\alpha=1/2$.}
\label{fig2}
\end{figure}

Figure \ref{fig2} depicts the phase difference given by Eq.~(\ref{argument}). One can check that the slope of $\delta(\mm)$ approaches a constant value, 
$$
\delta'(\mm)\to -\frac{2 \alpha \, \text{arctanh}(\alpha )}{1-\alpha^2},
$$
as $\mm\to\infty$.

\begin{figure}
\includegraphics[width=0.9\linewidth]{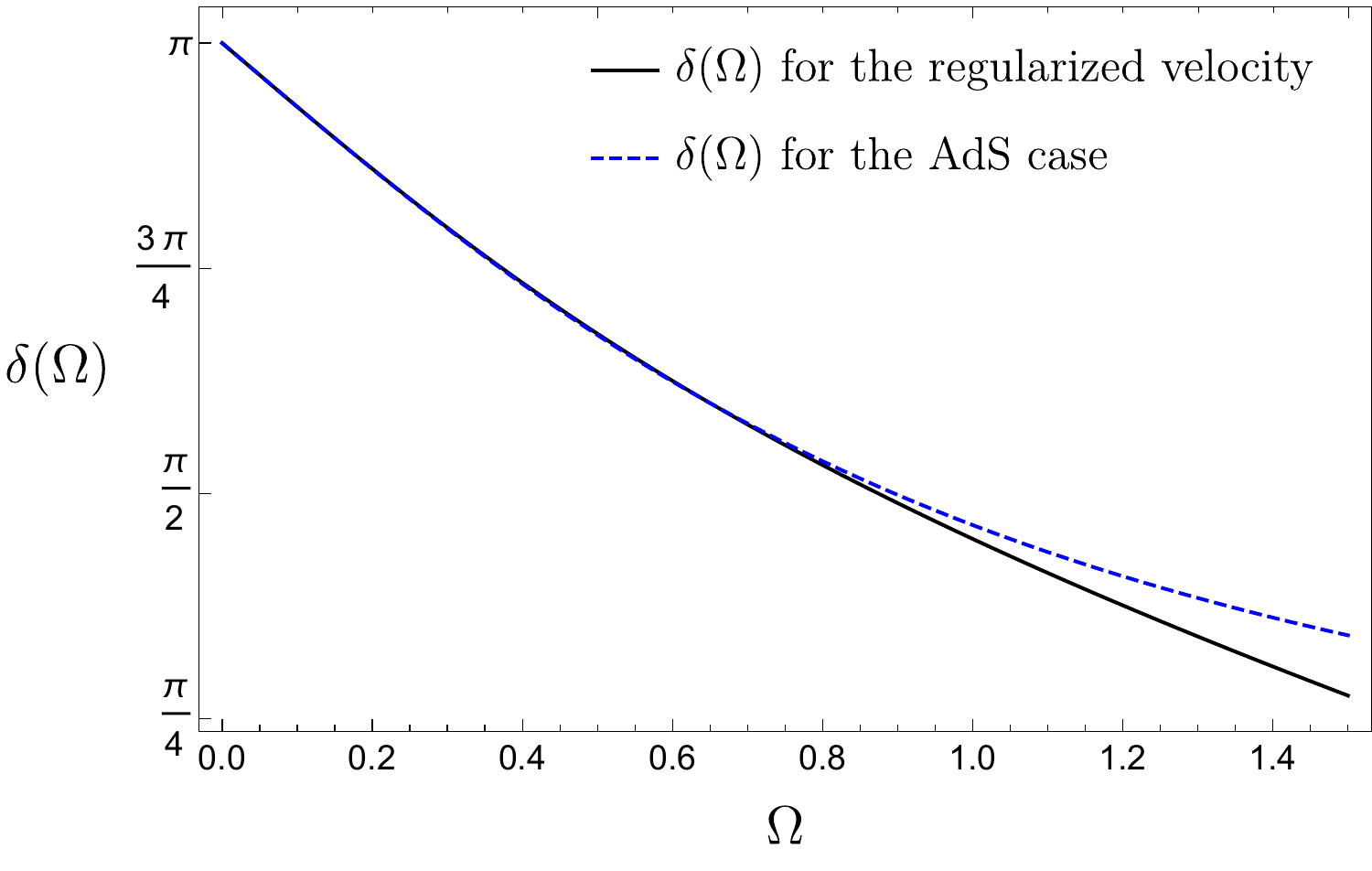}
\caption{Phase difference between ingoing and outgoing circular waves for the $\ads$ case (dashed blue) and for the regularization of $\vec{v}$ given by Eq. (\ref{tanh})  (solid black), with $\alpha=1/2$. Both curves agree for small values of $m$ when $\beta$ is chosen as in Eq. (\ref{betasol}).}
\label{fig3}
\end{figure}

One can also check that, for small values of $\mm$,
$$
\delta(\mm) \sim \pi -\frac{\mm}{1-\alpha^2} \ln{\left(\frac{4}{1-\alpha ^2}\right)}.
$$
Figure \ref{fig3} shows that this phase difference agrees with that obtained for the $\ads$ case for small $\mm$. In order to analyze this, let us go back to dimensionful quantities. The dimensionless frequency $\mm=\omega\rho_0/c$ can then be  written as
$$
	\mm=2\pi\rho_0/\lambda,
$$
where we used the fact that $\lambda c=2\pi\omega$, $\lambda$ being the wavelength of the sound wave. Recall that the parameter $\rho_0$ represents the region where $v(\rho)$ raises from $0$ to its constant value. For small values of $\mm$, the sound wavelength is much larger than $\rho_0$ and, therefore, the sound waves cannot properly probe the region for which $v(\rho)$ is not constant. As a consequence, the results found in \cite{anads} hold. 

Comparing the results coming from Eqs. (\ref{argument0}) and (\ref{argument}), we find perfect agreement up to first order in $\mm$ as long as
\begin{equation}
	\beta=\ln\left(\frac{2}{\sqrt{1-\alpha^2}}\right) \, \rho_0.
\label{betasol}
\end{equation}
This illustrates the fact, discussed in \cite{anads}, that the extra boundary condition in the $\ads$ case, determined by $\beta$, provides an effective description of the point source at $r=0$. The equation above corresponds to the value of $\beta$ associated with the regularization given by Eq. (\ref{tanh}). Changing the regularization also changes this effective parameter. 

We finally note that there are no damped modes for this fluid configuration. This follows from the fact that finiteness of energy requires that a solution of the wave equation for imaginary $\mm$ would have to be proportional to $R_1(x)$ for $x\to0$, and proportional to $R_{-}(x)$ for $x\to\infty$. This cannot happen since $R_{1}$ and $R_{-}$ are linearly independent.

\section{Conclusion}

We studied an analogue model based on radial flows in hydrodynamics. It is known that for constant radial velocities the resulting effective metric corresponds to the $\ads_2\times S^1$ spacetime \cite{anads}, which is not globally hyperbolic. This implies that the dynamics of fields in this background is not well defined unless extra boundary conditions are prescribed (in this case at the spatial boundary of $\ads$). On the analogue model end this implies that one needs to specify extra boundary conditions at the origin. This corresponds to an effective description of how the field interacts with the point source/sink of the flow. Here we considered regularizations of the fluid velocity near the source/sink at the origin. We found that a certain class of regularizations---those for which $\vec{v}(r)$ is smooth at the origin---leads to a well-defined dynamics for sound waves without the need of extra boundary conditions. This is to be expected since, at least as far as the velocity field is concerned, the hole at the origin has become invisible in those cases. On the effective spacetime end this corresponds to the introduction of a deformation of $\ads$ near its spatial infinity (so that the spacetime is forced to no longer be asymptotically $\ads$). We finally showed that, when the wavelength of the sound waves is much larger than the effective radius set by the regularization, the effects of the latter are negligible, as expected. In this case the regularization has the effect of picking up a specific boundary condition for the idealized case so that both models agree in the long wavelength limit.

\acknowledgments
It is a pleasure to acknowledge discussions with M. Richartz. The authors acknowledge support from FAPESP Grant No. 2013/09357-9. D. Q. A. acknowledges support from CAPES Grant No. 1490213/2015. R. A. M. acknowledges support from FAEPEX Grant No. 519292. J. P. M. P. acknowledges support from FAPESP Grant No. 2016/07057-6.


\begin{thebibliography}{99}
\bibitem{endnote1} By that we mean continuity of the vector field $\vec{r}\mapsto\vec{v}(\vec{r})$ as map from $\mathbb{R}^2$ to $\mathbb{R}^2$. This map is continuous at the origin only if $\vec{v}(\vec{0})=\vec{0}$. It is also easily seen that, if $v(r)\propto r^n$, differentiability of this vector field requires that $n\ge1$.
\bibitem{endnote2} We should note that a (very different) analogue model for $\ads$ was introduced in~\cite{campos} in the context of a Randall-Sundrum brane model, a $4+1$-dimensional spacetime which is asymptotically $\ads$ in  the bulk.
\end{thebibliography}
\end{document}